*Review*

# Filled Carbon Nanotubes as Anode Materials for Lithium-Ion Batteries

Elisa Thauer [1], Alexander Ottmann [1], Philip Schneider [1], Lucas Möller [1], Lukas Deeg [1], Rouven Zeus [1], Florian Wilhelmi [1], Lucas Schlestein [1], Christoph Neef [1], Rasha Ghunaim [2,3], Markus Gellesch [2], Christian Nowka [2], Maik Scholz [2], Marcel Haft [2], Sabine Wurmehl [2,4], Karolina Wenelska [5], Ewa Mijowska [5], Aakanksha Kapoor [6], Ashna Bajpai [6], Silke Hampel [2] and Rüdiger Klingeler [1,7,*]

[1] Kirchhoff Institute for Physics, Heidelberg University, INF 227, Heidelberg 69120, Germany; elisa.thauer@kip.uni-heidelberg.de (E.T.); alex.ottmann@posteo.de (A.O.); schneider_philip@web.de (P.S.); lucas.moeller@me.com (L.M.); lukas-deeg@gmx.de (L.D.); rouven.zeus@gmx.net (R.Z.); florianwilhelmi@gmx.de (F.W.); lucasschlestein@gmx.de (L.S.); Christoph.Neef@isi.fraunhofer.de (C.N.)

[2] Leibniz Institute for Solid State and Materials Research (IFW) Dresden, Dresden 01069, Germany; rgonaim@ppu.edu (R.G.); M.Gellesch@bham.ac.uk (M.G.); c.nowka@ifw-dresden.de (C.N.); maik.scholz@ifw-dresden.de (M.S.); m.haft@ifw-dresden.de (M.H.); s.wurmehl@ifw-dresden.de (S.W.); s.hampel@ifw-dresden.de (S.H.)

[3] Department of Applied Chemistry, Palestine Polytechnic University, Hebron P.O. Box 198, Palestinian Territories

[4] Institute for Physics of Solids, Technical University of Dresden, Dresden 01062, Germany

[5] Nanomaterials Physicochemistry Department, Faculty of Chemical Technology and Engineering, West Pomeranian University of Technology, Szczecin 71-065, Poland; Karolina.Wenelska@zut.edu.pl (K.W.); emijowska@zut.edu.pl (E.M.)

[6] Indian Institute of Science Education and Research, Pune 411 008, India; aakanksha.kapoor@students.iiserpune.ac.in (A.K.); ashna@iiserpune.ac.in (A.B.)

[7] Centre for Advanced Materials (CAM), Heidelberg University, INF 225, Heidelberg 69120, Germany

* Correspondence: klingeler@kip.uni-heidelberg.de

**Abstract:** Downsizing well-established materials to the nanoscale is a key route to novel functionalities, in particular if different functionalities are merged in hybrid nanomaterials. Hybrid carbon-based hierarchical nanostructures are particularly promising for electrochemical energy storage since they combine benefits of nanosize effects, enhanced electrical conductivity and integrity of bulk materials. We show that endohedral multiwalled carbon nanotubes (CNT) encapsulating high-capacity (here: conversion and alloying) electrode materials have a high potential for use in anode materials for lithium-ion batteries (LIB). There are two essential characteristics of filled CNT relevant for application in electrochemical energy storage: (1) rigid hollow cavities of the CNT provide upper limits for nanoparticles in their inner cavities which are both separated from the fillings of other CNT and protected against degradation. In particular, the CNT shells resist strong volume changes of encapsulates in response to electrochemical cycling, which in conventional conversion and alloying materials hinders application in energy storage devices. (2) Carbon mantles ensure electrical contact to the active material as they are unaffected by potential cracks of the encapsulate and form a stable conductive network in the electrode compound. Our studies confirm that encapsulates are electrochemically active and can achieve full theoretical reversible capacity. The results imply that encapsulating nanostructures inside CNT can provide a route to new high-performance nanocomposite anode materials for LIB.

## 1. Introduction

Lithium-ion batteries (LIB) offer high gravimetric and volumetric energy densities which renders them particularly suitable for mobile applications. In order to optimize their performance, in particular with larger energy density, there is a continuous search for novel electrode materials. Electrode materials based on conversion and alloying mechanisms promise extremely enhanced electrochemical capacities in lithium-ion batteries as compared to conventional materials [1–3]. However, severe fading of the electrochemical capacity due to fractionation, resulting from pronounced volume changes upon electrochemical cycling, is one of the major drawbacks with respect to application. In addition to volume changes associated with the conversion reaction, low electric conductivity of many conversion materials seriously hinders their applicability in secondary batteries [4]. Nanosizing promises enhanced capability to accommodate strain induced by electrochemical cycling and may reduce kinetic limitations of the macroscopic counterparts of electrode materials [5–7] since downsizing particles yields shorter diffusion lengths and hence enhances rate performances of electrode materials. However, low density limiting volumetric energy densities of actual electrodes as well as high surface areas are relevant issues to be considered in nanomaterials as well. High reactivity associated with high surface area typically promotes irreversible processes and associated electrolyte consumption. In this respect, due to carbon's restricted voltage regime of electrochemical activity, carbon (nano) coating is a valuable tool to protect active nanomaterials, thereby avoiding enhanced electrolyte degradation and associated (and potentially dangerous) gas production [8]. Downscaling materials towards carbon-shielded hybrid nanomaterials hence offers a route to obtain electrode materials for LIB with enhanced performance.

Rational design of electrode materials has to tackle the abovementioned issues of low electronic conductivity limiting many promising electrode materials as well as of large volume changes during electrochemical cycling, with the latter particularly causing electrode structure and particles distortions and hence strong performance fading. Hierarchical nanocomposite carbon/active material structures offer an effective way to solve these issues as such materials exploit size effects of the nanoscaled building blocks [9–14]. Mechanical strain arising from volume changes is additionally buffered by the hierarchical structures. In this way, such materials optimally maintain the integrity of the bulk material while offering improved electrical conductivity owing to a carbon-based backbone structure [15–28]. Moreover, a strong backbone structure improves the stability of the composite with respect to mechanical strain arising from volume changes during electrochemical cycling.

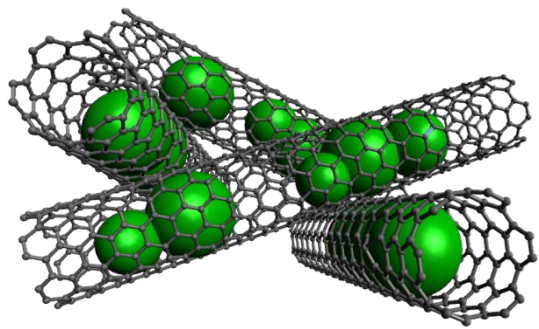

**Figure 1.** Schematics of nanocomposite material formed by interconnected carbon nanotubes (CNT) filled with high-capacity electrode materials. Essential characteristics are (1) size-controlled nanoparticles in the inner cavities of CNT which are separated from encapsulates in other CNT, (2) electrical contact of the incorporated material to a stable conductive network of CNT, (3) limitation of direct electrolyte/active material contact yielding and hence improved chemical stability. Created with Avogadro [29].

We report CNT-based composite nanomaterials with enhanced electrochemical performance realized by filling material into CNT (for a schematics see Fig. 1) which is electrochemically active when nanoscaled [30]. CNT display excellent conductivity as well as mechanical and chemical stability which renders them an excellent carbon source in hybrid nanomaterials [31]. However, in conventional approaches using exohedrally functionalized CNT, synthesis of uniformly sized and shape-controlled nanoparticles is challenging. In addition, while the interconnected network of carbon nanotubes provides an electrically conducting backbone structure, decorated nanoparticles onto the outer CNT-walls tend to lose electrical contact upon cycling-induced disintegration and particular methods have to be developed to improve connection to CNT [32–35]. Our results demonstrate successful synthesis of hybrid nanomaterial of CNT filled with $Mn_3O_4$, $CoFe_2O_4$, $Fe_xO_y$, Sn, and CoSn and show the electrochemical activity of encapsulated materials. Encapsulates are either conversion or alloying electrode materials which perform the following general reactions upon electrochemical cycling, respectively [2,36,37]:

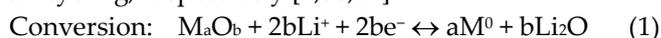

Conversion: $M_aO_b + 2bLi^+ + 2be^- \leftrightarrow aM^0 + bLi_2O$ (1)

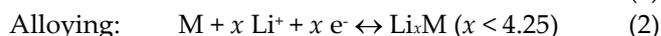

Alloying: $M + x\,Li^+ + x\,e^- \leftrightarrow Li_xM$ ($x < 4.25$) (2)

In this work, we demonstrate that in the case of conversion materials filled inside CNT, the encapsulated material completely participates in electrochemical cycling, i.e., the theoretical capacity is fully accessible. The backbone network of CNT is indeed unaffected by cracks of encapsulate which usually inhibit long-term stability. Our data hence imply that endohedrally functionalized CNT offer a promising route to new nanohybrid anode materials for LIB.

## 2. Synthesis and Characterization of Filled CNT

We report studies on hybrid nanomaterial of multiwalled carbon nanotubes (CNT) filled with $Mn_3O_4$, $CoFe_2O_4$, $Fe_xO_y$, Sn, and CoSn which have been fabricated by a variety of methods. Mostly, CNT of type PR-24-XT-HHT (Pyrograf Products, Inc., Cedarville OH, USA) have been used as templates. For introducing materials into the inner cavity of the CNT, mainly extensions of solution-based approaches reported in [38–43] have been applied [44,45]. This is illustrated by the example of $Mn_3O_4$@CNT which has been obtained by filling CNT with a manganese salt solution and a subsequent reducing step yielding homogeneously MnO-filled CNT (MnO@CNT) [4]. Subsequent heat treatment of MnO@CNT yields the complete conversion into $Mn_3O_4$@CNT, as confirmed by the XRD pattern in Figure 2. In case of filling with Co-Fe spinels, nitrate solutions of $Fe(NO_3)_3 \cdot 9H_2O$ (grade: ACS 99.0%–100.2%) and $Co(NO_3)_2 \cdot 6H_2O$ (grade: ACS 98.0%–102.0% metal basis) were used in stoichiometric ratios with respect to the metal ions (i.e., Fe:Co = 2:1). After adding CNT and treating the mixture in an ultrasonic bath with appropriate washing steps, the solid residue was dried and afterwards calcinated under argon flow atmosphere (100 sccm) at a temperature of 500 °C for 4 h to convert the nitrates into the corresponding cobalt ferrite. This is confirmed by XRD data in Figure 2 which indicate the presence of $CoFe_2O_4$. Pronounced peak broadening indicates the presence of nano-sized $CoFe_2O_4$ crystallites, with an estimated grain size of 20(5) nm by means of the Scherrer equation applied to the Bragg peak at 41.5°.

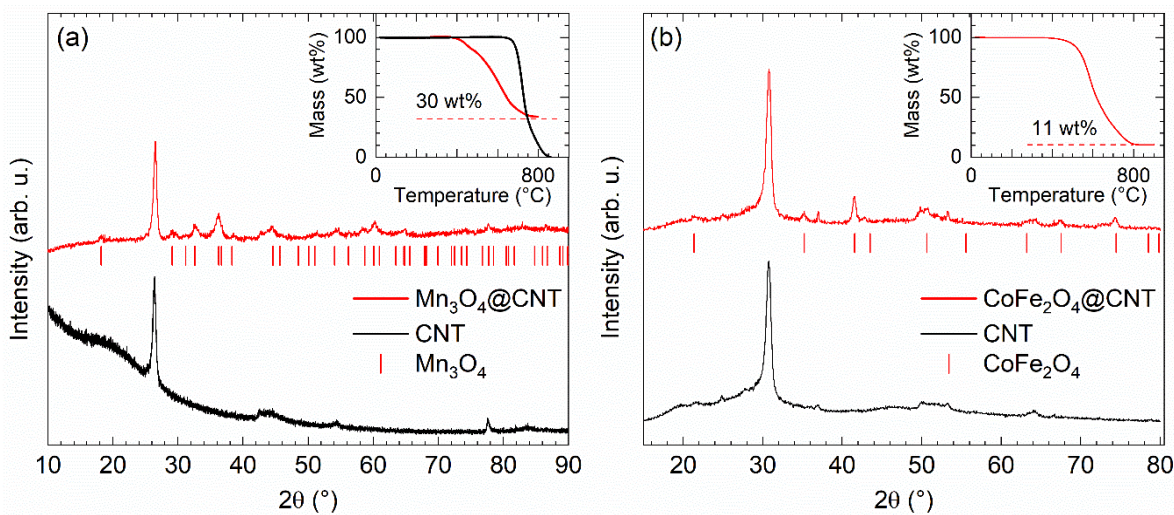

**Figure 2.** Left (**a**): PXRD patterns of Mn$_3$O$_4$@CNT and pure CNT. Vertical lines show the Bragg positions of Mn$_3$O$_4$ (space group I4$_1$/amd) [46]. Inset: Thermogravimetric analysis (TGA) data of Mn$_3$O$_4$@CNT and pure CNT. Right (**b**): PXRD patterns of CoFe$_2$O$_4$@CNT and of pristine CNT. Vertical ticks label Bragg positions of bulk CoFe$_2$O$_4$ (space group Fd$\bar{3}$m) [47]. Inset: TGA of CoFe$_2$O$_4$@CNT.

XRD patterns show relatively broad Bragg reflections which indicate small primary particle size of the noncarbon materials of the composite as expected for nanoparticles fitting inside the interior of CNT. This is confirmed by exemplary SEM and TEM studies presented in Figure 3. The images clearly show that the metal oxide nanoparticles are rather spherical and are located inside the CNT. Note the exception of possible nanowire formation in the case of metal-filled Sn@CNT as discussed in Section 3.4 (see Figure 15). The filling rate of Mn$_3$O$_4$@CNT is about 30(1) wt% and that of CoFe$_2$O$_4$@CNT (see the inset of Figure 2) is about 11(1) wt% as determined by thermogravimetric measurements (TGA).

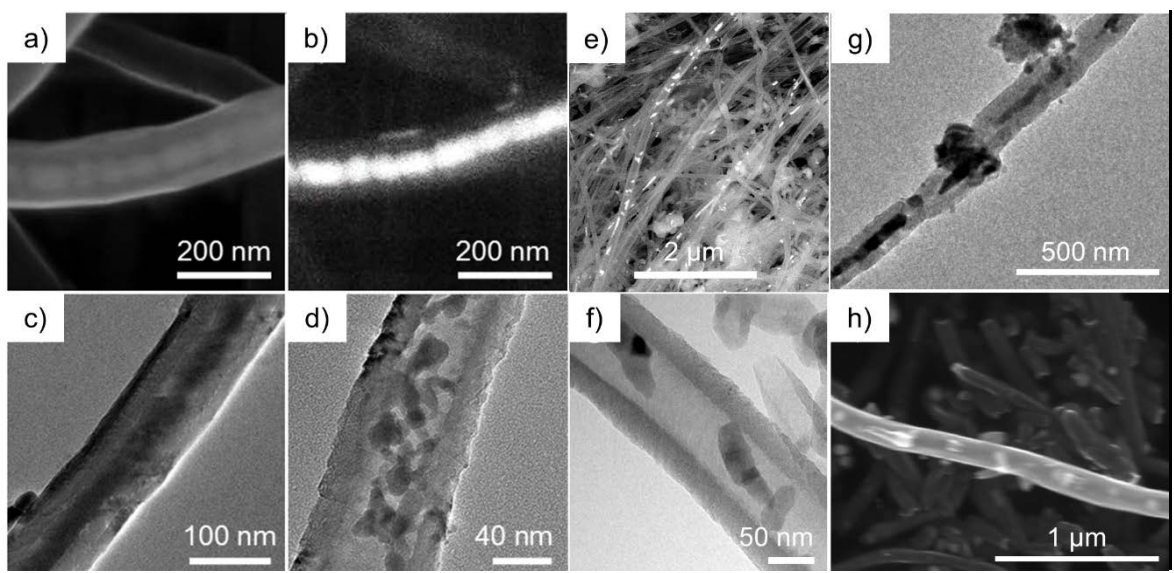

**Figure 3.** (**a**) SEM image of an individual Mn$_3$O$_4$@CNT (SE mode); (**b**) corresponding BSE mode image; (**c**,**d**) TEM images of different individual Mn$_3$O$_4$@CNT. Taken from [45]. (**e**) Overview SEM image of CoFe$_2$O$_4$@CNT (BSE mode); (**f**) TEM image of an individual CoFe$_2$O$_4$@CNT. (**g**) TEM image of an individual Fe$_x$O$_y$@CNT [48]. (**h**) SEM image of CoSn@CNT [49].

Electron microscopy confirms that the filling materials are located mainly inside the CNT. Exemplary SEM and TEM images are shown in Figure 3 (see also Figure 15 for Sn-filled CNT). In Mn$_3$O$_4$@CNT, the encapsulated particles are rather spherical with the average diameter of 15 ± 7 nm obtained by TEM analysis. Note, that this is smaller than the size-limiting inner diameter of the utilized CNT (~35 nm). The SEM overview image (Figure 3e) on CoFe$_2$O$_4$@CNT also confirms that the filling material is distributed along the inner cavity of the hollow CNT. TEM indicates spherical encapsulates as well as short rods inside CNT (Figure 3 e,f). Fe$_x$O$_y$@CNT (synthesis reported in [48]) appears to be mainly filled with α-Fe$_2$O$_3$ but also exhibits Fe$_3$O$_4$ as shown, e.g., by associated features in the magnetic susceptibility (see Section 3.3). Figure 3g also shows the presence of Fe$_x$O$_y$ nanoparticles outside CNT. In addition to separated spherical nanoparticles, encapsulates in CoSn@CNT and Sn@CNT form also nanowires up to 1 μm length (see Figure 3h and Figure 15). In either case, the encapsulates fill the complete inner diameter of the CNT, which is about 50 nm [44]. In summary, the results show that our synthesis approaches result in CNT filled with nanoparticles whose diameters are limited by the inner diameter of the CNT.

## 3. Electrochemical Studies

### 3.1. Mn$_3$O$_4$@CNT [30, 45]

Cyclic[1] voltammetry studies on Mn$_3$O$_4$@CNT and on pristine CNT, performed in the voltage range of 0.01–3.0 V vs. Li$^{0/+}$ and recorded at a scan rate of 0.1 mV s$^{-1}$, confirm electrochemical activity of encapsulates (Figure 4). During the initial cycle, starting with the cathodic scan, five distinct reduction peaks (R1–R5) and three oxidation peaks (O1–O3) are observed. The redox pair R1/O1 around 0.1 V and the irreversible reduction peak R3 at 0.7 V can be attributed to processes related to multiwalled CNT (Figure 4a). The irreversible reaction peak R3 signals formation of the solid electrolyte interphase (SEI) expected for carbon-based (here: CNT) systems [50]. The pronounced redox pair R1/O1 demonstrates that the bare CNT subsystem in the hybrid material is electrochemically active as it signals (de)lithiation of Li$^+$ ions between the layers of CNT [51,52]. Slight splitting of oxidation peak O1 indicates a staging phenomenon reported for graphite electrodes [37], and very similar behavior upon cycling is found in bare CNT [45]. All other features observed in Figure 4b are ascribed to the electrochemical reaction mechanism which has been reported for Mn$_3$O$_4$ as follows [53,54] (for further details see [45]):

(A) $\text{Mn}_3(\tfrac{1}{3}\cdot\text{II},\tfrac{2}{3}\cdot\text{III})\text{O}_4 + \text{Li}^+ + e^- \rightarrow \text{LiMn}_3(\tfrac{2}{3}\cdot\text{II},\tfrac{1}{3}\cdot\text{III})\text{O}_4$

(B) $\text{LiMn}_3\text{O}_4 + \text{Li}^+ + e^- \rightarrow \text{Li}_2\text{O} + 3\cdot\text{Mn(II)O}$

(C) $\text{Mn(II)O} + 2\cdot\text{Li}^+ + 2\cdot e^- \leftrightarrow \text{Li}_2\text{O} + \text{Mn}(0)$

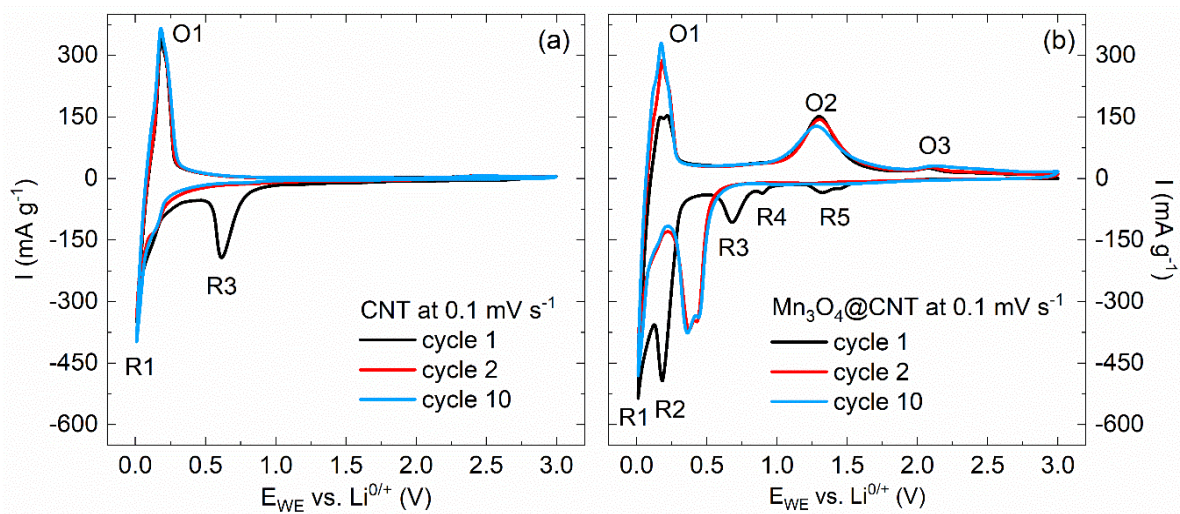

**Figure 4.** Cyclic voltammograms of (**a**) pristine CNT and (**b**) Mn$_3$O$_4$@CNT at 0.1 mV s$^{-1}$ [30].

The cyclic voltammograms (CVs) confirm electrochemical activity of encapsulated Mn$_3$O$_4$. Absence of significant changes between cycles 2 and 10 indicate good cycling stability which will be investigated in more detail below. Since the materials associated with the mechanism detailed in Equations (A) to (C) exhibit strong differences in magnetic properties, magnetic studies are suitable to follow the redox reaction. In particular, there are strong changes of magnetic properties upon electrochemical cycling from ferrimagnetic Mn$_3$O$_4$ to antiferromagnetic MnO (Figure 5; for further magnetization data see [45]). Pristine Mn$_3$O$_4$@CNT shows ferrimagnetic order below $T_C$ = 42 K as indicated by the magnetization data. In contrast, materials extracted after step (B) of the abovementioned redox reactions, i.e., after galvanostatic reduction at 5 mA g$^{-1}$ down to 0.5 V and passing the reduction peaks R5, R4, and R3 labelled in Figure 4b, displays nearly no traces of ferrimagnetic material. Quantitatively, the magnetization data indicate about 1% remainder of ferrimagnetic Mn$_3$O$_4$ after the first half cycle. Meanwhile, antiferromagnetic order is found below a temperature of ~120 K, which is expected for MnO [55] and is in agreement with Equation (B). Hence, our magnetometry data confirm electrochemical reactions as postulated in Equations (A–C) by tracking down individual magnetic species.

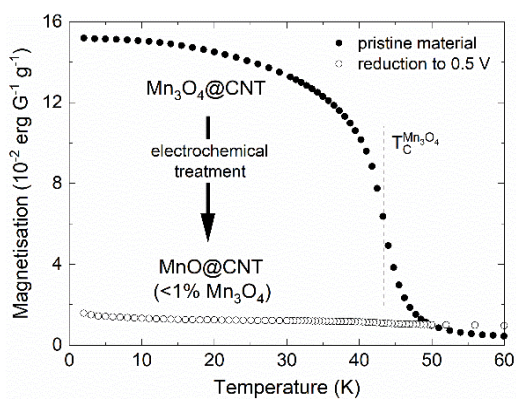

**Figure 5.** Magnetization of pristine and electrochemically cycled Mn$_3$O$_4$@CNT measured at B = 0.1 T (FC). The vertical line indicates the ferrimagnetic ordering temperature in Mn$_3$O$_4$.

Charge and discharge studies at specific current rates (Figure 6) display plateau-like regions in the voltage profiles signaling the redox features discussed above by means of Figure 4. In the initial cycle performed at 50 mA g$^{-1}$, specific charge and discharge capacities of 677 and 455 mAh g$^{-1}$, respectively, are achieved. Increasing the charge/discharge current to 100 and 250 mA g$^{-1}$, respectively, does not significantly affect the shape of the curves but yields smaller discharge capacities, e.g., 331 mAh g$^{-1}$ after 30 cycles. For higher currents, the plateaus corresponding to delithiation and lithiation of CNT vanish, while the conversion reaction (Equation (C)) is still visible in the data. The rate capability studies presented in Figure 6 display pronounced capacity losses when increasing charge/discharge currents. Specifically, maximum discharge capacities of 468, 439, 349, 245, and 148 mAh g$^{-1}$ are reached at 50, 100, 250, 500 and 1000 mA g$^{-1}$, respectively.

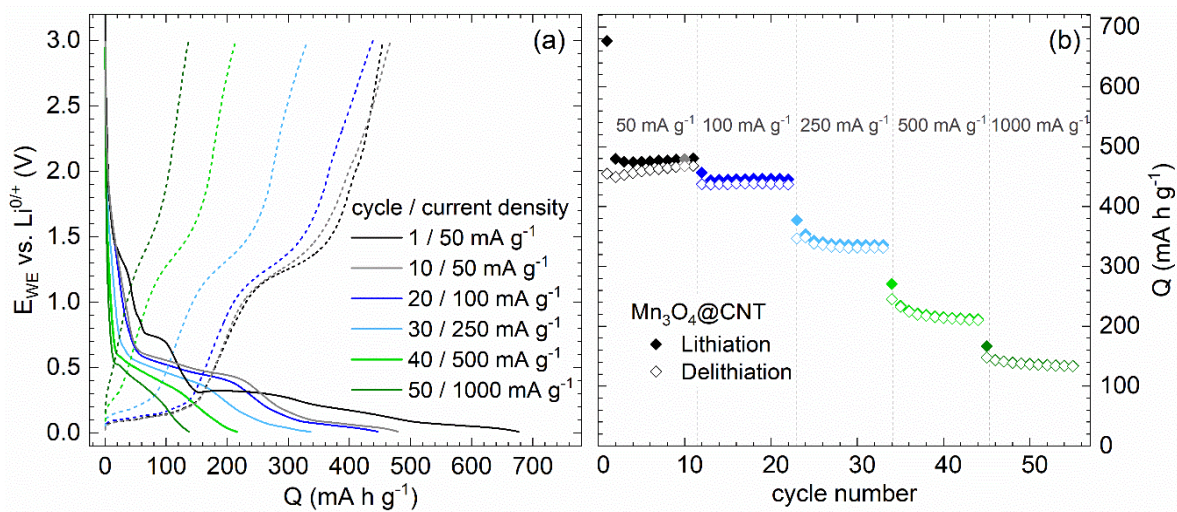

**Figure 6.** Rate capability studies of Mn3O4@CNT at 50, 100, 250, 500, and 1000 mA g$^{-1}$. (**a**) Potential profiles of specific lithiation (solid lines) and delithiation cycles (dashed lines). (**b**) Specific charge/discharge capacities upon cycling [30].

In order to assess the electrochemical performance of the composite with particular emphasis on the encapsulate, evolution of capacities at 100 mA g$^{-1}$ (galvanostatic cycling with potential limitation) upon cycling of Mn$_3$O$_4$@CNT and pristine CNT is shown in Figure 7. While the initial half cycle is strongly affected by irreversible processes associated with solid electrolyte interface (SEI) formation, the Mn$_3$O$_4$@CNT nanocomposite exhibits increasing capacities for approximately 15 cycles in contrast to decreasing values of pristine CNT. The nanocomposite reveals a maximum discharge capacity of 463 mA h g$^{-1}$ in cycle 18, of which 93% is maintained after 50 cycles (429 mA h g$^{-1}$). Thus, incorporation of Mn$_3$O$_4$ into CNT leads to more than 40% enhanced specific capacities on average as compared to unfilled CNT. The data, i.e., on filled and unfilled CNT, enable calculating the specific capacity of incorporated Mn$_3$O$_4$ (29.5 wt%). The encapsulate's initial capacity of about 700 mAh g$^{-1}$ increases significantly to 829 and 820 mAh g$^{-1}$ (cycle 18) and declines thereafter, with capacity retention of around 90% after 50 cycles. The Mn$_3$O$_4$ capacity even exceeds the theoretical expectations of the conversion reaction (C) from cycle 6 on (dashed line in Figure 7). This might be associated with a capacity contribution due to oxidative feature O3 (Figure 4b), which supposedly indicates the back-formation of Mn$_3$O$_4$ and corresponding reduction processes [56,57]. Note, however, the error bars of 5% due to mass determination of encapsulate and subtraction of data on pristine CNT. Initial capacity increase was also observed in previous studies on Mn$_3$O$_4$/CNT composites [58,59].

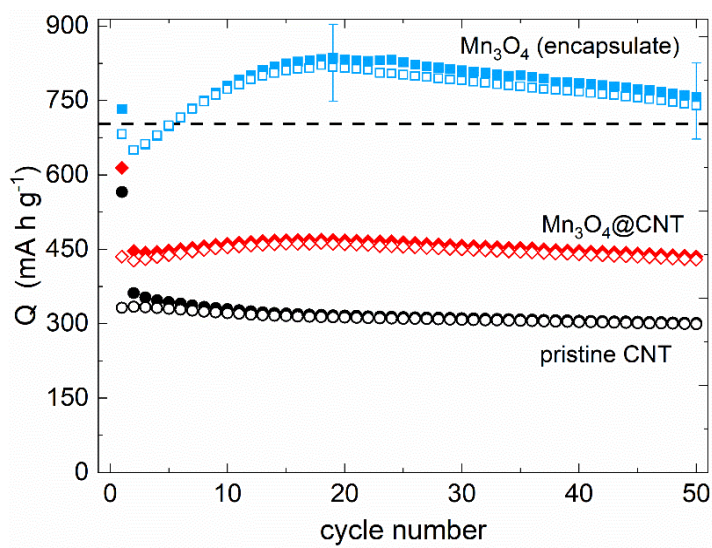

**Figure 7.** Specific charge/discharge capacities at 100 mA g$^{-1}$ of pristine CNT, Mn$_3$O$_4$@CNT, and calculated capacity of the encapsulate. The dashed line shows the theoretical capacity of the reversible conversion reaction (C) [30].

Our analysis shows that full conversion between MnO and metallic Mn can be achieved reversibly and the maximum of the contributed capacity by the Mn$_3$O$_4$ encapsulate is accessible (Figure 7). In particular, the nanoparticles inside CNT are completely involved in the electrochemical processes. This finding is supported by the fact that the active material inside CNT experiences distinct structural changes, as evidenced by TEM studies (Figure 8). Figure 8 b,c presents materials after 13 galvanostatic cycles, at 100 mA g$^{-1}$, taken after delithiation and lithiation. No clear differences are observed between the lithiated and the subsequently delithiated material. In both cycled materials, the encapsulate which initially exhibits well-defined, rather spherical nanoparticles has developed extended patches. The TEM image also shows lower contrast of the encapsulate to the CNT environment which is indicative of lower density of the encapsulate. Equations (A) and (B) indeed suggest rather larger volume expansion of Mn$_3$O$_4$ during initial lithiation and concomitant agglomeration as well as amorphization of the filling which is in agreement with the TEM results. Notably, despite the strong changes of encapsulate, CNT mantles still display the characteristic graphitic layers of multiwalled carbon nanotubes (see Figure 8d). Hence, electrochemical cycling does not severely damage the structure of the CNT. Furthermore, an amorphous layer of ~5 nm thickness can be observed on top of the graphitic CNT layers, which can be attributed to the SEI. The TEM analysis hence shows that the CNT indeed offer a stable environment for the manganese oxides which is able to accommodate the strain due to volume expansion during electrochemical cycling and guarantees a consistent electrical contact to the active material.

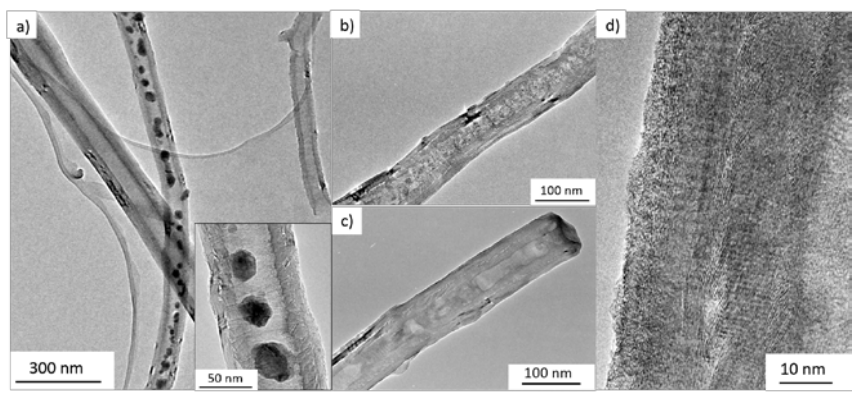

**Figure 8.** TEM images of (**a**) uncycled, (**b**) galvanostatically lithiated, and (**c**) delithiated Mn$_3$O$_4$@CNT. (**d**) High-resolution TEM image of a CNT shell of delithiated material after 13 cycles. Taken from [45].

*3.2. CoFe$_2$O$_4$@CNT*

As shown in Figure 3 f,g, nanosized particles of cobalt ferrite CoFe$_2$O$_4$ are incorporated into CNT by a similar procedure as applied in the case of Mn$_3$O$_4$@CNT. The mass content of CoFe$_2$O$_4$ in the composite materials however amounts to only 11 wt%, leading to smaller effects of the encapsulate. In order to evaluate the benefits of CNT shells, the electrochemical performance of the nanocomposite CoFe$_2$O$_4$@CNT is compared to that of bare CoFe$_2$O$_4$ nanoparticles (Figure 9). In general, electrochemical lithium storage of up to 8 Li$^+$/f.u. in CoFe$_2$O$_4$ follows a conversion mechanism (Equation (D)), which may be preceded by initial intercalation of Li$^+$ ions into the original ferrite structure [60]:

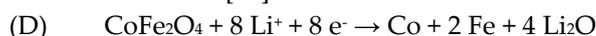
(D)    CoFe$_2$O$_4$ + 8 Li$^+$ + 8 e$^-$ → Co + 2 Fe + 4 Li$_2$O
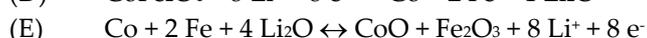
(E)    Co + 2 Fe + 4 Li$_2$O ↔ CoO + Fe$_2$O$_3$ + 8 Li$^+$ + 8 e$^-$

Both processes show up as redox features in the CVs in Figure 9 which for CoFe$_2$O$_4$@CNT also show features present in pristine CNT (Figure 4a) [45,51,61,62].

In bare CoFe$_2$O$_4$ nanoparticles (Figure 9a), the initial half cycle reduction peaks indicate, at 1.5 V, initial intercalation into the spinel structure (R0), and at 1.1 and 0.55 V indicate R1/SEI formation. In addition, there is a shoulder at 0.95 V and a peak at 0.01 V (R2). In all subsequent reductive half cycles, the most pronounced reduction peak occurs at 0.85 V (R1*). Expectedly, R0 vanishes after the first cycle. The oxidative scans display a broad oxidation double peak between 1.5 V and 2.5 V with a maximum intensity around 1.65 V (O1). R1 most likely indicates both conversion of the spinel to Co and Fe [60,63] and SEI formation [64], while R2 signals intercalation of Li$^+$ ions into added carbon black [64,65]. Upon further cycling, Co and Fe oxidize to CoO and Fe$_2$O$_3$, respectively (O1), followed by the corresponding conversion processes at R1* (Equation (D)) [36,63,66–69].

CVs on CoFe$_2$O$_4$@CNT in Figure 9b show features associated with CoFe$_2$O$_4$ superimposed by redox peaks related to CNT. In the initial cycle, features attributed to CoFe$_2$O$_4$ appear at 1.6 (R0), 1.2, and 0.7 V (SEI) with a shoulder at 0.8 V (R1). Upon further cycling, they are shifted to 1.6 (R0) and 0.9 V (R1*). Reversible oxidation peaks appear at similar voltages as compared to bare CoFe$_2$O$_4$ nanoparticles, i.e., between 1.5 and 2.0 V with a maximum at 1.55 V (O1). The results imply smaller overpotentials in CoFe$_2$O$_4$@CNT as compared to the bare CoFe$_2$O$_4$ nanoparticles, indicating improved energy efficiency. Furthermore, cycling stability is superior, yielding noticeable redox activity of the CoFe$_2$O$_4$ encapsulate in the 10th cycle. Both improvements can be attributed to benefits of the CoFe$_2$O$_4$@CNT composite material, i.e., to enhanced overall conductivity and better structural integrity.

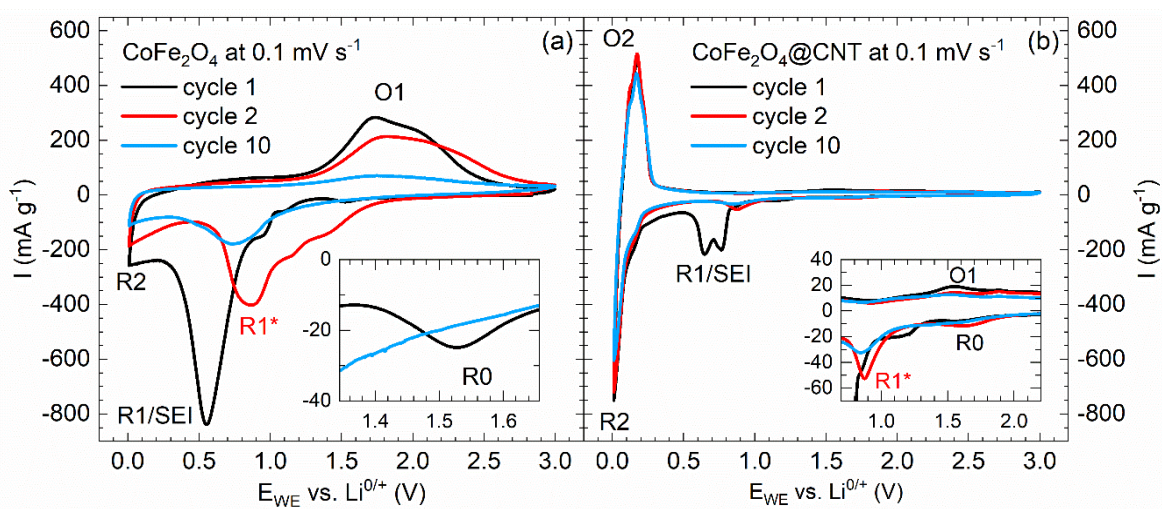

**Figure 9.** Cyclic voltammograms of (**a**) pristine CoFe$_2$O$_4$ and (**b**) CoFe$_2$O$_4$@CNT, at 0.1 mV s$^{-1}$ [30].

These conclusions are corroborated by galvanostatic cycling with potential limitation (GCPL) data (Figure 10). Firstly, higher capacities of CoFe$_2$O$_4$@CNT as compared to pristine CNT imply electrochemical activity of encapsulates for 60 cycles under study. In addition to irreversible effects associated with SEI formation, there are capacity losses, in particular in initial cycles, so that the electrode demonstrates only 97% of Coulombic efficiency after 15 cycles. Capacity retention of CoFe$_2$O$_4$@CNT amounts to a fair value of 76% after 60 cycles (243 mAh g$^{-1}$). Analogously to Section 3.1, the specific contribution of CoFe$_2$O$_4$ is evaluated by subtracting the measured capacities of pristine CNT, weighted with the mass ratio of 89:11 (CNT:CoFe$_2$O$_4$). The analysis shows (Figure 10b) that both for pristine and CNT-encapsulated CoFe$_2$O$_4$ there are pronounced capacity losses upon cycling while the initial capacities exceed the theoretical maximum value of 914 mAh g$^{-1}$ due to SEI formation. CNT-encapsulated active material clearly outperforms bare CoFe$_2$O$_4$ nanoparticles. To be specific, after 20 cycles, 475 mAh g$^{-1}$ (71%) is retained in CoFe$_2$O$_4$@CNT while the bare particles show 190 mAh g$^{-1}$ (22%). This result again demonstrates that embedding nanosized CoFe$_2$O$_4$ inside CNT partly compensates for the typical capacity fading associated with the conversion reactions upon electrochemical delithiation or lithiation known for spinel materials.

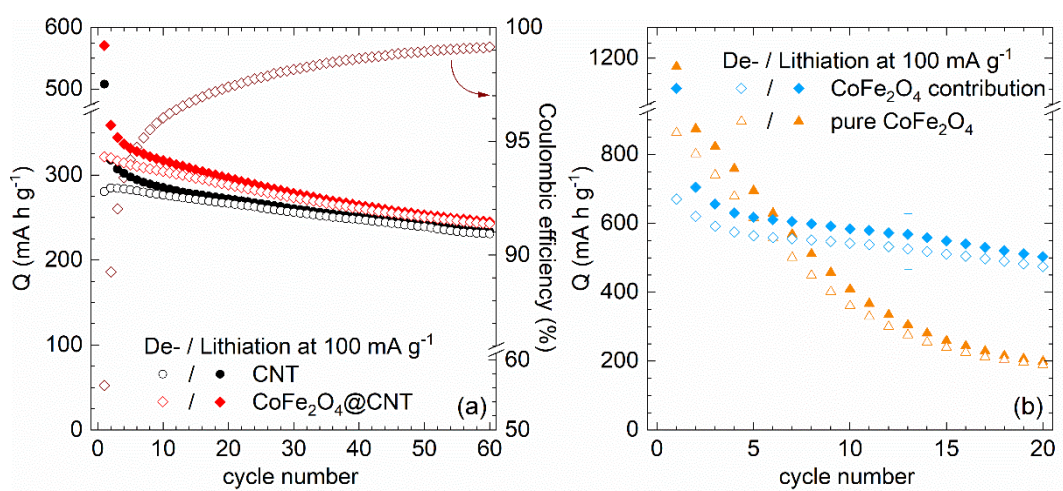

**Figure 10.** (**a**) Specific charge/discharge capacities, at 100 mA g$^{-1}$, of pristine CNT and CoFe$_2$O$_4$@CNT as well as the Coulombic efficiencies of the latter. (**b**) Capacity contribution of the encapsulated CoFe$_2$O$_4$ in comparison to pristine CoFe$_2$O$_4$ [29].

While encapsulated CoFe$_2$O$_4$@CNT demonstrates electrochemical activity, it is illustrative to compare the results with alternative carbon/CoFe$_2$O$_4$ hybrid nanomaterials. Direct comparison is often hindered by the fact that the carbon-related capacity is not always subtracted as done here. For many carbon/CoFe$_2$O$_4$ hybrid materials, much higher values than maximum theoretical capacity of CoFe$_2$O$_4$ are reported. A value of 1046 mAh g$^{-1}$ is reported for mesoporous CoFe$_2$O$_4$ nanospheres cross-linked by carbon nanotubes [70]. Porous carbon nanotubes decorated with nanosized cobalt ferrite show 1077 mAh g$^{-1}$, after 100 cycles [69]. More than 700 mAh g$^{-1}$ of total capacity of the composite was obtained when CoFe$_2$O$_4$ is encapsulated into carbon nanofibers with 36% carbon content [71]. A list of recently achieved record values may be found in [72]. We note that excessive capacity beyond the theory values in transition metal oxide/carbon nanomaterials have been associated, e.g., to decomposition of electrolyte and formation of a polymer/gel-like film on the nanoparticles [73]. Another hypothesis refers to interface charging effects by lithium accommodation at the metal/Li$_2$O interface [74]. Our data indeed suggest that surface effects might be relevant as CNT-encapsulation of active material evidently suppresses this phenomenon.

*3.3. Fe$_x$O$_y$@CNT and CNT@Co$_3$O$_4$*

Fe$_x$O$_y$@CNT has been synthesized as described in [48]. XRD and magnetic characterization studies [30] imply the presence of several iron oxides (i.e., of α-Fe$_2$O$_3$ as well as of γ-Fe$_2$O$_3$ or/and Fe$_3$O$_4$) in the materials. While the main phase appears as α-Fe$_2$O$_3$, magnetic studies show both the Morin and Verwey transitions which enable to unambiguously identify α-Fe$_2$O$_3$ and Fe$_3$O$_4$, respectively. Note, that the presence of antiferromagnetic γ-Fe$_2$O$_3$ can neither be confirmed nor excluded by our magnetic studies. Analyzing the magnetization data indicates the presence of ferromagnetic iron oxide (i.e., γ-Fe$_2$O$_3$ and/or Fe$_3$O$_4$) of about 30(8) wt%.

The CVs shown in Figure 11a display two reductions (R1, R2) and two oxidations (O1, O2) which are observed in all cycles. We attribute R1/O1 to electrochemical activity of CNT. Except for typical initial irreversible effects at R2/SEI, all features are well explained by electrochemical processes known in iron oxides. Mechanisms in α-Fe$_2$O$_3$ as identified by Larcher et al. [75,76] involve Li-intercalation in nanoparticles, followed by conversion to metallic Fe and Li$_2$O via intermediately formed cubic Li$_2$Fe$_2$O$_3$. This process is partly reversible as it includes formation of FeO [77] and γ-Fe$_2$O$_3$ [78,79]. For Fe$_3$O$_4$, after initial intercalation, Li$_2$Fe$_3$O$_4$ is formed which is subsequently reduced to Fe and Li$_2$O [80,81]. In all iron oxides present in Fe$_x$O$_y$@CNT, including γ-Fe$_2$O$_3$, electrochemical processes display similar features which are not well distinguishable [82,83]. The inset of Figure 11a presents a weak reduction peak R3 which we attribute to abovementioned Li-intercalation into iron oxides. Note, that the second peak in the inset is due to an intrinsic cell setup effect. Conversion reactions appear at around 0.6 V and are signaled by feature R2. The shoulder at 0.8 V indicates the successive nature of the lithiation processes. Upon cycling, R2 shifts to 0.9–1.2 V, thereby indicating significant structural changes due to the initial conversion process. The large width of O2 might indicate several oxidation processes upon delithiation. The evolution of the oxidation features upon cycling implies severe fading effects.

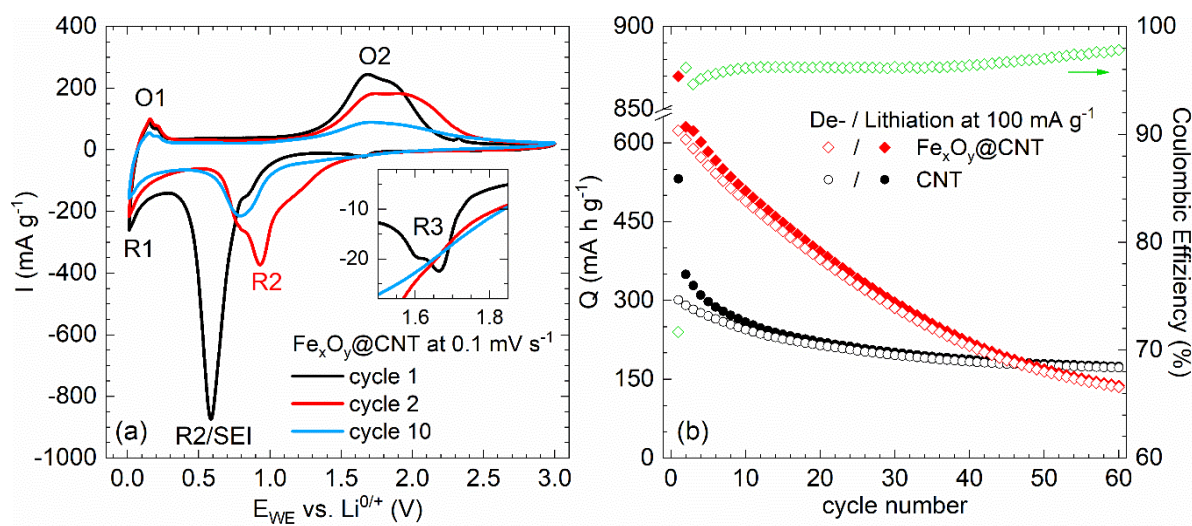

**Figure 11.** (**a**) Cyclic voltammogram of Fe$_x$O$_y$@CNT at 0.1 mV s$^{-1}$. (**b**) Specific charge/discharge capacities, at 100 mA g$^{-1}$, of pristine CNT (Pyrograf Products, type PR-24-XT-HHT) and Fe$_x$O$_y$@CNT, as well as the Coulombic efficiencies of the latter [30].

This is confirmed by the data in Figure 11b which presents specific charge/discharge capacities of Fe$_x$O$_y$@CNT obtained at 100 mA g$^{-1}$. Respective data on bare CNT (Pyrograf Products, type PR-24-XT-HHT) are shown for comparison. The initial capacities of the composite amount to 870 and 624 mAh g$^{-1}$, which reflects initial irreversible processes. There is a strong decay in capacity which yields only 78% (489 mAh g$^{-1}$) in cycle 10 and 26% (165 mAh g$^{-1}$) in cycle 50 of the initial discharge capacity. The results clearly show that envisaged improvement of cycling stability due to encapsulation into CNT is not achieved. Presumably, iron oxide content outside CNT is rather large so that a significant part of functionalization is exohedral. In such case, we assume that volume changes upon cycling leads to detachment of these particles from the CNT network which results in diminished electrochemical activity. In contrast, [84] reports α-Fe$_2$O$_3$-filled CNT which show 90% capacity retention in cycle 50.

Inferior stability of exohedrally functionalized CNT upon electrochemical cycling is further confirmed, e.g., for CNT decorated by mesoporous cobalt oxide (CNT@Co$_3$O$_4$). The material was synthesized as reported in [34]. The composite exhibits 41 wt% of mesoporous Co$_3$O$_4$ spheres with mean diameters between 100 and 250 nm decorated to the CNT network. The electrochemical behavior of CNT@Co$_3$O$_4$ (Figure 12a) during the initial cycle shows SEI formation and the initial reduction process of Co$_3$O$_4$ to metallic cobalt and formation of amorphous Li$_2$O during the initial cycle. Double peaks appearing in cycle 2 correspond to a multistep redox reaction caused by the Co$^{2+}$/Co$^0$ and Co$^{3+}$/Co$^{2+}$ couples [85,86]. The integrated specific capacities calculated from the CVs (Figure 12b) display significant capacity fading upon continued cycling. For comparison, a blend of separately fabricated CNT and Co$_3$O$_4$ nanoparticles were mechanically mixed postsynthesis in the same ratio of 59% CNT and 41% Co$_3$O$_4$ which, according to TGA, is realized in the decorated CNT@Co$_3$O$_4$ nanocomposite. The blend's CV shows similar peak positions as found in CNT@Co$_3$O$_4$, and a similarly high reductive capacity is measured for the postsynthesis blend in the first cycle. However, the associated reversible capacity is much lower as compared to the CNT@Co$_3$O$_4$ hybrid nanomaterial and the irreversible loss between charge and discharge capacity is higher. After a few cycles, both the blend and CNT@Co$_3$O$_4$ show similarly low performance, which indicates that the benefit of attaching mesoporous Co$_3$O$_4$ to the surface of CNT has completely faded, presumably due to detachment of the mesoporous Co$_3$O$_4$ nanospheres [34].

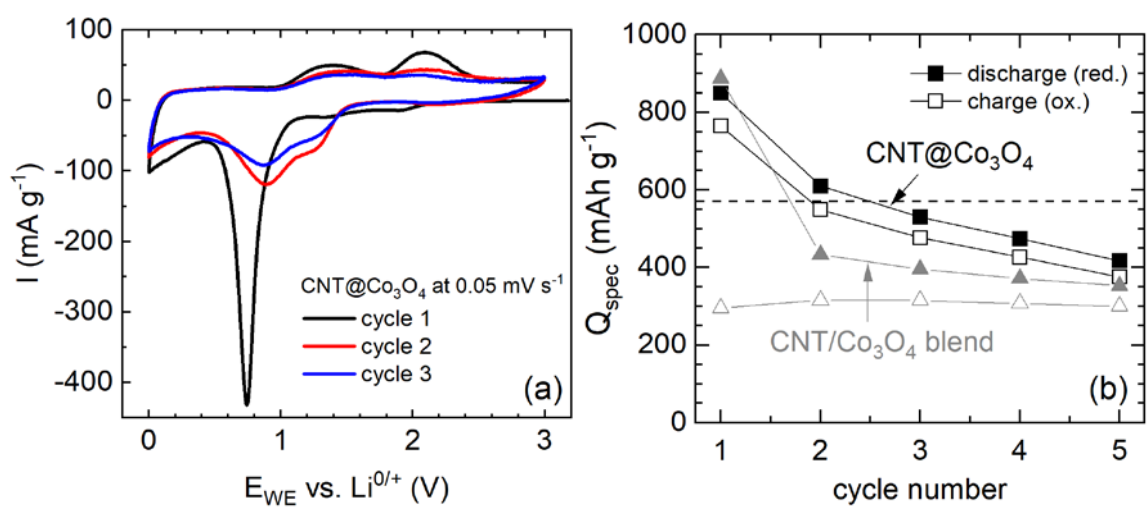

**Figure 12.** (**a**) CV curves of CNT@Co$_3$O$_4$, at 0.05 mV s$^{-1}$ in the voltage range of 0.01–3.00 V. Integrated charge and discharge capacities for five cycles as deduced from CV [32].

*3.4. Sn@CNT and CoSn@CNT*

The alloying process described by Equation (2) implies feasibility of (semi)metallic electrode materials for electrochemical energy storage. Using M = Ge, Sn, the alloy Li$_x$M is formed with x up to 4.25 Li$^+$/f.u. [87,88]. While Ge exhibits lower molecular weight and good Li$^+$-diffusivity, Sn is much cheaper and exhibits higher electrical conductivity [89]. For Sn, the most Li-rich alloy is Li$_{17}$Sn$_4$ (= Li$_{4.25}$Sn) which implies a theoretical capacity of 960 mAh g$^{-1}$ [90]. Upon lithiation, several stable alloys such as LiSn and Li$_7$Sn$_2$ are formed, resulting in complex (de)alloying processes of several stages which are associated with large volume changes [91]. In CoSn@CNT, Co is electrochemically inactive and is supposed to buffer the volume changes as similarly done in a commercial Sn-Co-C composite by Sony [87,92,93].

Synthesis of Sn@CNT has been published in [44]. While the encapsulate in Sn@CNT is β-Sn with a filling ratio of 20 wt%, encapsulate in CoSn@CNT is a mixture of β-Sn, CoSn, and mainly CoSn$_2$ with in total 17 wt% of Sn and 5 wt% of Co. In addition to encapsulated separated spherical nanoparticles, encapsulates in Sn@CNT also form nanowires up to 1 μm in length. Both spheres and wires fill the complete inner diameter of the CNT, which is about 50 nm [44].

The CVs of Sn@CNT- and CoSn@CNT-based electrode materials shown in Figure 13 are similar to each other and confirm the multistage processes expected from reports on non-CNT materials. In both cases, in addition to the SEI formation, the peaks R1/O1 signal electrochemical activity of CNT. The reduction peak R3 at 0.6 V and the pair R2/O2 at 0.3 V as well as several features at 0.35–0.85 V, are all attributed to multi-stage (de)alloying processes.

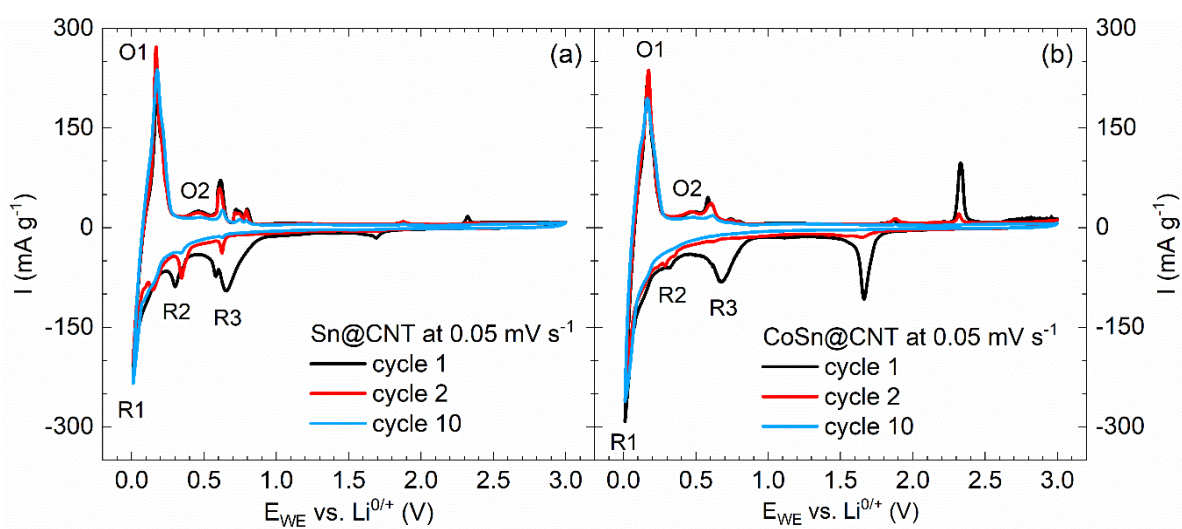

**Figure 13.** CVs of Sn@CNT (a) and CoSn@CNT (b) in the regime 0.01–3.0V vs. Li/Li+ at a scan rate 0.05 mV/s. Note that the oxidation peaks at 1.9 and 2.3 V and the reduction peak at 1.7 V appearing in the first two cycles are due to the experimental cell setup [30].

Galvanostatic cycling of Sn@CNT and CoSn@CNT as compared to pristine CNT (Pyrograf Products, Typ PR-24-XT-HHT) quantifies the contribution of encapsulates to the materials' capacities (Figure 14a). Sn@CNT displays clearly improved values. Quantitatively, the data imply an initial reversible capacity of 322 mAh g$^{-1}$ in cycle 2, of which 281 mAh g$^{-1}$, i.e., 87%, is retained in cycle 50. In contrast, fading is much more severe in CoSn@CNT, which shows only 66% retained of the initial capacity 317 mAh g$^{-1}$, i.e., its performance in cycle 50 falls below that of pristine CNT. As will be discussed below, these data show that there is no positive (buffering) effect of alloyed Co. Rate capacities shown in Figure 14b at cycling rates 100–2000 mA g$^{-1}$ illustrate the strong effect of fast cycling on both materials, thereby confirming limiting kinetics of the underlying electrochemical alloying processes.

Figure 14a also presents the specific capacity of the Sn encapsulate which is derived by correcting the data by the effect of pristine CNT. Note error bars of Sn-capacity of up to 20% resulting in particular from errors in determining the filling ratio. In the first cycle, the reversible capacity amounts to 589 mAh g$^{-1}$ which suggests deintercalation of x = 2.6 Li$^+$/f.u. Capacity fading is about 15% between cycle 5 (495 mAh g$^{-1}$) and cycle 50 (422 mAh g$^{-1}$). Even the initial capacities are much smaller than the theoretical one of 960 mAh g$^{-1}$ that would be achieved for x = 4.25.

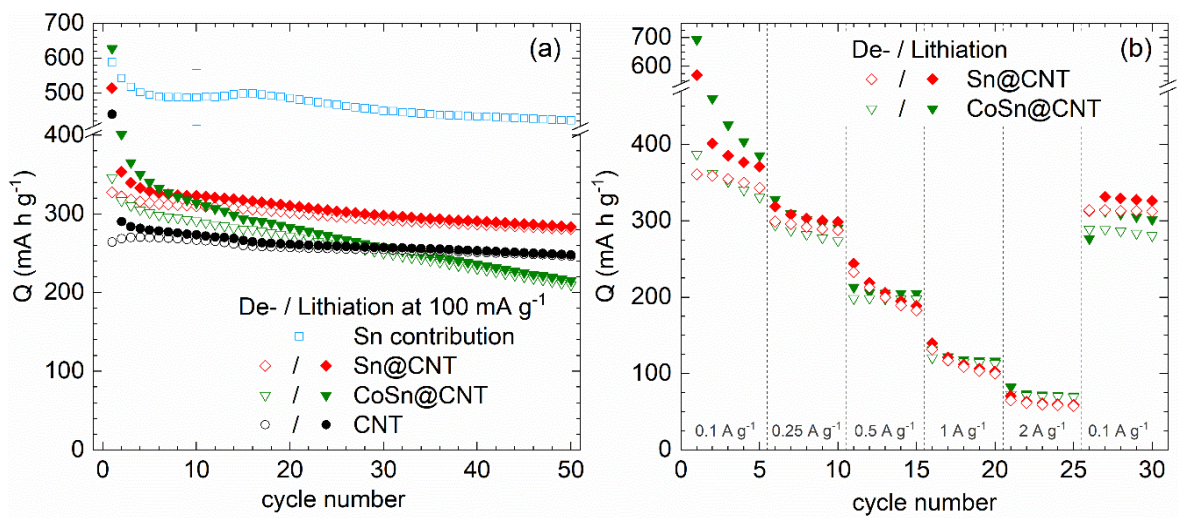

**Figure 14.** Specific capacities of Sn@CNT-, CoSn@CNT-, and pristine CNT-based electrodes. (**a**) Galvanostatic cycling with potential limitation (GCPL) at 100 mA g$^{-1}$. Blue data markers show the specific capacity of the Sn encapsulate after correcting the contribution of CNT. (**b**) GCPL at different rates of 0.1–2.0 A g$^{-1}$ [30].

In-situ XRD studies on $Li_xSn_y$ have shown that intermediate phases $Li_2Sn_5$ and LiSn are expected [94]. In agreement with these studies, the presence of (at least) two reduction peaks in the CVs of both materials (see Figure 13) suggests at least a two-stage process in the materials at hand. However, comparison to the literature does not allow to attribute these peaks to a specific process. This also holds for the observed (at least) four oxidation peaks which indicate step-wise delithiation of the $Li_xSn_y$-alloy. For CoSn@CNT where the encapsulate mainly consists of $CoSn_2$, Mössbauer studies have shown the formation of $Li_xSn$ with x ≈ 3.5 in the first cycle [95]. Such a process is not visible in the CV (Figure 13b) but the respective feature might be masked by the SEI-peak. It is argued in [95] that, upon delithiation, $Li_{-3.5}Sn$ forms an amorphous $Li_xCo_ySn_2$-matrix which is crucial for the expected buffering associated with Co-alloying. We conclude that, in CoSn@CNT, this $Li_xCo_ySn_2$-matrix is not realized but Co just deteriorates the electrochemical performance. This conclusion is supported by the fact that the CVs in Figure 13 display the same number of peaks at similar potentials in both Sn@CNT and CoSn@CNT, which indicates identical processes. We assume separation of Co and Sn instead of $Li_xCo_ySn_2$-formation yielding electrochemically inactive regions.

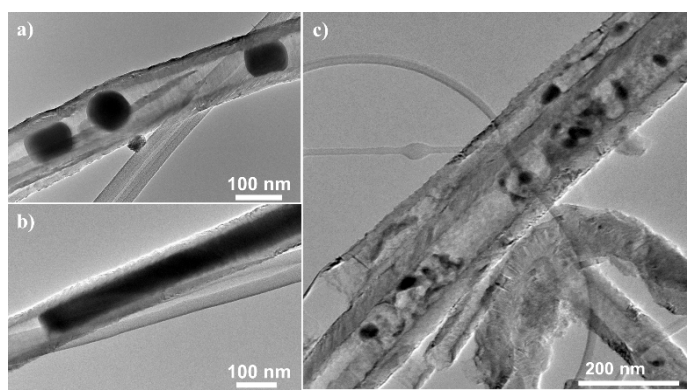

**Figure 15.** TEM images of pristine (**a**,**b**) and galvanostatically cycled (10 cycles) Sn@CNT (**c**).

The effect of galvanostatic cycling, at 50 mA g$^{-1}$, on Sn@CNT is demonstrated by TEM images in Figure 15. In the cycled materials, well separated homogenous encapsulates (Figure 15a,b) in the pristine material convert to rather completely but inhomogeneously filled CNT whose filling is indicated by different TEM contrast, i.e., different densities of encapsulate. These finding agrees with expected volume changes, in particular to large expansion upon lithiation, and phase separation of encapsulated material. One may speculate that the dark regions visible after cycling indicate electrochemically-inactive domains of Sn. The presence of inactive regions would be in agreement to the GCPL data (Figure 14) which show that only a maximum of 60% of the full Sn-capacity is achieved. Finally we note that previous studies on Sn-filled CNT have demonstrated better performance as compared to the material at hand. Wang et al. have reported Sn@CNT with filling ratios of 38 wt% and 87 wt% [96]. The former, i.e., less filled, CNT have demonstrated superior performance with capacities of 500 mAh $g^{-1}$ for 80 cycles at 100 mA $g^{-1}$. The relevant parameter seems to be the size of Sn particles, which was 6−10 nm in [96]. Larger encapsulates filling the complete inner diameter of the CNT of about 50 nm as realized in the materials at hand seem to be detrimental and may cause electrochemically inactive regions. Addition of Co as a potential buffer does not improve the performance of such rather large nanoparticles inside CNT but even causes additional capacity fading.

**4. Experimental Methods**

*4.1. Material Characterization*

Materials were characterized by X-ray diffraction (XRD) with either Stadi P (Stoe, Darmstadt, Germany) using Cu K$\alpha$1 radiation ($\lambda$ = 1.5406 Å) or X'Pert Pro MPD PW3040/60 (PANanalytical, Almelo, Niederlande) using Co K$\alpha$ radiation ($\lambda$ = 1.79278 Å). Thermogravimetric analysis (TGA) was carried out with SDT Q600 (TA Instruments, Waters Corporation, Milford, Massachusetts, U.S.City, Abbr of State if USA, Country). The morphology was investigated by means of scanning electron microscopy (SEM, Nova NanoSEM 200 (FEI company, Hilsboro, Oregon, U.S.)) and transmission electron microscopy (TEM, JEM-2010F (JEOL, Akishima, Japan), Tecnai (FEI company, Hilsboro, Oregon, U.S.)). A MPMS-XL5 (Quantum Design, San Diego, Californis, U.S.) superconducting quantum interference device (SQUID) magnetometer was used to perform magnetic measurements.

*4.2. Electrochemical measurements*

Electrochemical properties were studied by cyclic voltammetry (CV) and galvanostatic cycling with potential limitation (GCPL) in Swagelok-type cells [97]. The measurements were performed on a VMP3 potentiostat (BioLogic) at a temperature of 25 °C. For the preparation of the working electrode, the active material was optionally mixed with carbon black (Super C65, Ymeris Graphite and Carbon , Bironico, Switzerland , City, Abbr of State if USA, Country) and stirred in a solution of polyvinylidene fluoride (PVDF, Solvay , Brussels, Belgium company, City, Abbr of State if USA, Country) in N-methyl-2-pyrrolidone (NMP) for at least 12 h. After evaporat,ing most of the NMP in a vacuum oven (80 °C, <10 mbar) the spreadable slurry was applied on copper mesh current collectors (Ø 10 mm). The as-prepared electrodes were dried at 80 °C in a vacuum oven (<10 mbar), mechanically pressed at 10 MPa, and afterwards dried again. The assembly of cells was done in a glovebox under argon atmosphere ($O_2/H_2O$ < 5 ppm) using a lithium metal foil disk (Alfa Aesar, Haverhill, Massachusetts, U.S. City, Abbr of State if USA, Country) pressed on a nickel current collector as counter electrode. The electrodes were separated by two layers of glass microfibre (Whatman GF/D) soaked with 200 µL of a 1 M $LiPF_6$ salt solution in 1:1 ethylene carbonate and dimethyl carbonate (Merck ElectrolyteLP30). For post cycling studies, working electrodes were washed three times in dimethyl carbonate and afterwards dried under vacuum.

**5. Conclusions**

Endohedral functionalization of multiwalled carbon nanotubes by means of high-capacity electrode materials is studied with respect to application for electrochemical energy storage. Encapsulation indeed yields size-controlled nanoparticles inside CNT. The presented data imply that the filled materials are electrochemically active and can achieve full theoretical reversible capacity. While conversion and alloying processes yield cracks and amorphization of the encapsulate, the CNT mantles are found to be only very little affected by electrochemical cycling. The backbone network of CNT hence maintains its integrity and improved performance with respect to unshielded or exohedrally-attached nanomaterials. For appropriately tailored materials, CNT-based nanocomposites show smaller overpotentials and hence improved energy efficiency as well as improved cycling stability. The results imply that encapsulating nanostructures inside CNT provides a successful route to new high-performance nanocomposite anode materials for LIB.

**Author Contributions:** Conceptualization, E.T., A.O. and R.K.; methodology, E.T., A.O., S.H., A.B., E.M. and R.K.; data analysis, E.T., A.O. and R.K.; synthesis and material characterization, R.G., M.G., C.N. (Christian Nowka), M.S., M.H., S.W., S.H., A.K., A.B., K.W. and E.M.; electrochemical investigations, E.T., A.O., P.S., L.M., L.D., R.Z., F.W., L.S. and C.N. (Christoph Neef); writing—original draft preparation, E.T., A.O., and R.K.; writing—review, E.T., A.B., E.M., C.N., A.O, S.H., S.W. and R.K.

**Funding:** Work was partly supported by Deutsche Forschungsgemeinschaft DFG via KL 1824/12-1 and the CleanTech-Initiative of the Baden-Württemberg-Stiftung (Project CT3: Nanostorage). KW and EM acknowledge financial support from the National Science Center Poland (UMO-2016/23/G/ST5/04200).

**Acknowledgments:** The authors thank G. Kreutzer for technical support.

**Conflicts of Interest:** The authors declare no conflict of interest.

[1]This subchapter in parts reviews Refs. [30] and [45].